# Tip-induced nitrene generation

Leonard-Alexander Lieske[1,*], Aaron H. Oechsle[1,#], Igor Rončević[2], Ilias Gazizullin[3], Florian Albrecht[1], Matthias Krinninger[4,†], Leonhard Grill[3], Friedrich Esch[4], Leo Gross[1,*]

[1] IBM Research Europe – Zurich, Rüschlikon, Switzerland
[2] Department of Chemistry, University of Manchester, Manchester, United Kingdom
[3] Physical Chemistry Department, University of Graz, Graz, Austria
[4] Chair of Physical Chemistry and Catalysis Research Center, Department of Chemistry, TUM School of Natural Sciences, Technical University of Munich, Garching, Germany

[#] Present address: Laboratory for X-ray Nanoscience and Technologies, Center for Photon Science, Paul Scherrer Institute, 5232-Villigen, Switzerland
[†] Present address: Leibniz Supercomputing Centre (LRZ) of the Bavarian Academy of Sciences and Humanities, Boltzmannstr. 1, 85748 Garching, Germany
*Corresponding author, email: LAL@zurich.ibm.com, LGR@zurich.ibm.com

**Abstract:**
We generated trinitreno-*s*-heptazine, a small molecule featuring three nitrene centers, by tip-induced chemistry from the precursor 2,5,8-triazido-*s*-heptazine on bilayer NaCl on Au(111). The precursor's azide groups were dissociated to form mono-, di- and trinitreno-*s*-heptazine, yielding molecules with one to three nitrene centers. The precursor and its products are characterized by atomic force microscopy and scanning tunnelling microscopy. Broken-symmetry DFT and configuration interaction calculations of inter- and intra-nitrene exchange couplings suggest a ferromagnetic coupling of the $S = 1$ nitrene centers, resulting in a high-spin septet ground state for neutral trinitreno-*s*-heptazine in the gas phase. On bilayer NaCl on Au(111), the combined results of experiments and theory suggest trinitreno-*s*-heptazine to be an anion with a sextet ground state.

**Introduction:**
Nitrenes are known as highly reactive intermediates present in many reactions of nitrogen-containing compounds.[1–3] Because of their high reactivity, nitrenes are studied enclosed in crystals or matrices,[4–7] but are generally difficult to isolate.[8] In this study, the tip-induced generation of individual nitrene sites in individual molecules is realized using 2,5,8-triazido-*s*-heptazine[9–13] as the precursor. Denitrogenation of triazido-*s*-heptazine has been previously achieved by X-ray illumination,[14] photocatalysis,[14] and thermal decomposition.[11,13,14] Heptazine-[9,10,15–17] and triazine-based[18–23] structures are of interest since a long time[24] and are promising candidates for bottom-up growth of nitrogen-rich covalent organic frameworks based on carbon-nitride-based structures and 2D films.[11,25–27]

Individual nitrene centers are generally considered to be $sp^1$-hybridized, with high-spin (triplet) ground states.[8,28–31] Electron spin resonance performed on nitrene-containing molecules showed mononitrenes in a triplet ground state, dinitrenes in a quintet ground state and trinitrenes in a septet ground state.[31–34] Such molecules exhibit a strong localization of unpaired electrons on the nitrene centers.[1,35] This strong localization is attributed to the small mixing of the heteroatom orbitals with the π-conjugated system.[36] The on-surface tip-induced generation of nitrenes might prove useful for the investigation and coupling of molecules with multiple localized spin centers.

Non-contact atomic force microscopy (AFM) with CO-functionalized tips,[37] allows for resolving the structure of individual molecules. Additionally, scanning tunneling microscopy (STM) can be used to gain information about the open-/closed-shell character and spin ground states of molecules.[38–41]

Atom manipulation is a useful tool to perform redox chemistry[41–45] and to dissociate constituent groups from precursors.[46–49]

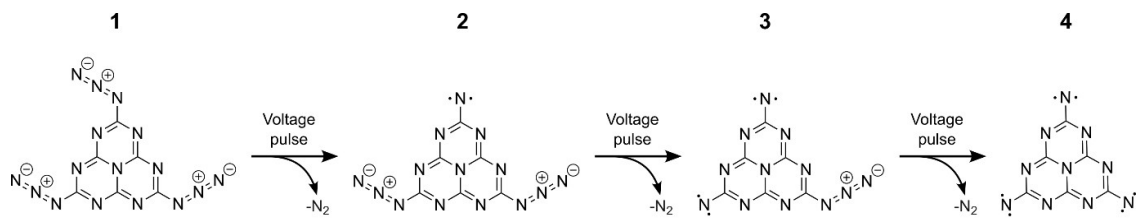

**Figure 1. Structures of triazido-*s*-heptazine and its tip-induced products** representing the reaction sequence to form a molecule with three nitrene centers. From 2,5,8-triazido-*s*-heptazine (TAH), **1**, shown in a zwitterionic resonance structure, mononitreno-*s*-heptazine **2**, dinitreno-*s*-heptazine **3** and trinitreno-*s*-heptazine **4** can be generated by voltage pulses using the tip of the STM/AFM.

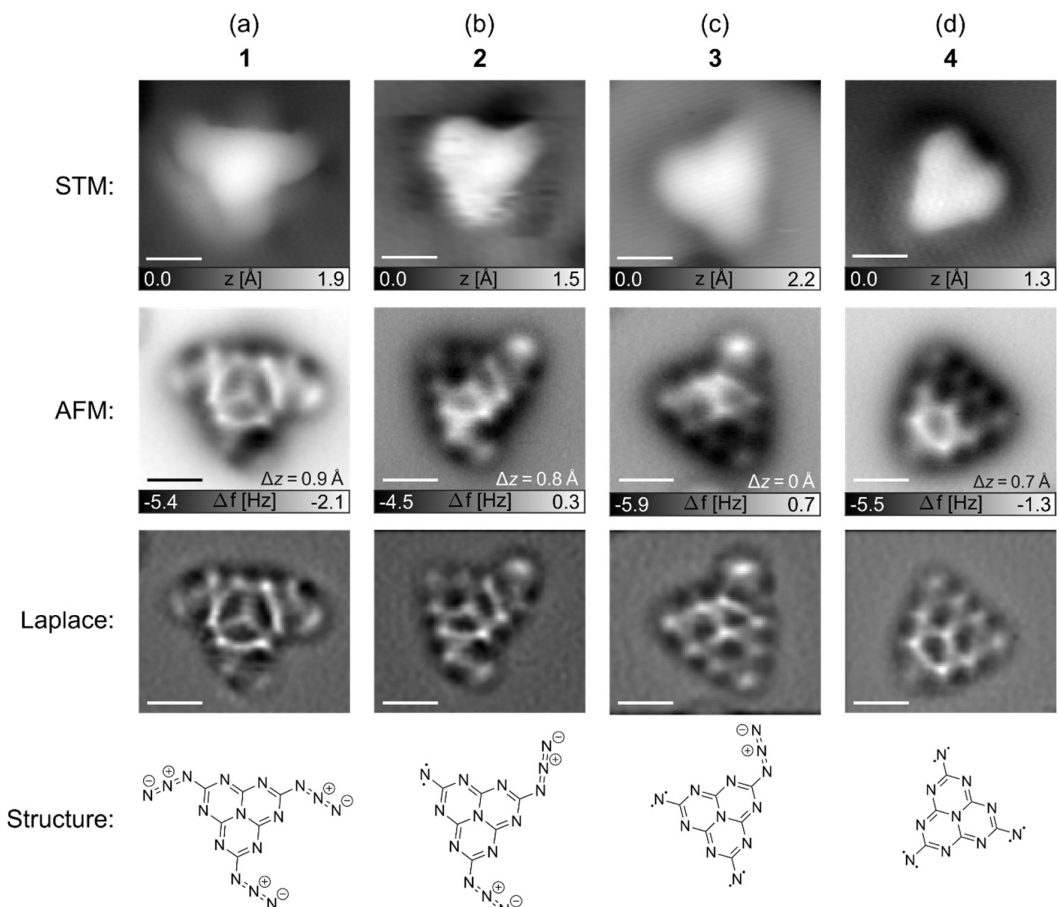

**Figure 2. STM and AFM measurements of 1, 2, 3 and 4 on NaCl(2ML)/Au(111). (a)** TAH **1**, **(b)** mononitrene **2**, **(c)** dinitrene **3** and **(d)** trinitrene **4**. Constant-current STM measurements are shown in the first row, constant-height AFM measurements in the second row and corresponding Laplace-filtered AFM images in the third row. For each molecule, a correspondingly oriented resonance structure is shown in the fourth row. STM parameters in (a), (b), (d) *I* = 1 pA, *V* = 0.2 V, in (c) *I* = 0.5 pA, *V* = 0.5 V. For AFM, tip-height offsets Δ*z* from the respective STM parameters as setpoints are indicated. A positive Δ*z* corresponds to an increase in tip-sample distance from the setpoint. Scalebars are 5 Å.

**Results and discussion:**

To obtain and characterize molecules that contain one, two or three nitrene groups, we generated them using tip-induced chemistry from the precursor 2,5,8-triazido-*s*-heptazine **1** (TAH).[14,50,51] The reaction scheme is shown in Figure 1.

Nitrene generation by azide-group dissociation, i.e., cleaving off $N_2$, was performed by applying voltage pulses with the tip placed above a molecule. The applied pulse voltages $V_P$ ranged from $V_P$ = 2 V to 2.6 V resulting in maximum tunnelling currents $I_P$ in the range of $I_P$ = 0.2 pA to 20 pA during a pulse (see Supplementary Note 1 for a detailed description of the voltage pulses for nitrene generation). Increased values of $V_P$ and $I_P$ were needed for each subsequent azide-group dissociation from **1** to **2** to **3**. Sequential tip-induced dissociation of azide groups starting from **1** to form **2**, **3** and **4** is shown in Fig. S2.

Constant-current STM and constant-height AFM measurements of **1** on bilayer NaCl are shown in Figure 2a. STM shows a central protrusion at the location of the heptazine core surrounded by three fainter features located at the positions of the azide groups. AFM resolves the heptazine core atomically. We observe a brighter contrast (more positive Δ*f* signal) on the nitrogen atoms compared to the carbon atoms, in agreement with computational studies of *s*-triazine moieties.[52,53] At the small tip-sample distances in our experiment, tilting of the CO molecule at the tip enhances the imaging contrast but is known to lead to apparent image distortions,[37,54–59] which result here in distorted six-membered rings of the heptazine core.[55,60,61] The presence of nitrogen atoms in heterocycles has been shown to cause apparent distortions in AFM images.[62,63] In addition, the contrast in AFM images provides information about the molecular adsorption geometry.[64]

The AFM data on **1** (Fig. 2a) indicate that the heptazine core of **1** is planar and adsorbed parallel to the surface. The orientations of all azide groups can be inferred from the AFM contrast. Figure S1 shows the experimentally determined and the simulated adsorption position of **1**. The AFM simulation agrees with the experimental AFM data (Fig. S1c and Fig. S1d). In the simulated geometry, we find that each azide group is oriented toward a sodium cation of the NaCl surface (Figure S1e).

In the STM measurement of mononitreno-*s*-heptazine **2** (mononitrene) (Figure 2b) the nitrene moiety appears with a more circular shape compared to the appearance of an azide group. In the AFM measurement, the nitrene center, i.e., the remaining N atom of the dissociated azide group, is atomically resolved. Additionally, the AFM data suggests a tilted heptazine core of **2** with respect to the NaCl surface. That is, at the position of the generated nitrene (top, left corner of the molecule in Figure 2b) the AFM contrast of the heptazine core appears darker indicating that the molecule is closer to the surface at this location. In dinitreno-*s*-heptazine **3** (dinitrene) (Figure 2c) AFM resolves the two nitrene centers and again indicates a tilt of the heptazine core with respect to the sample surface. In the AFM measurement of trinitreno-*s*-heptazine **4** (trinitrene) (Figure 2d) all three nitrene centers are resolved and the heptazine core appears adsorbed with a small tilt angle with two nitrene centers close to the surface (top, and right corner of the molecule in Figure 2d) and one further away from the surface (bottom left corner of the molecule in Figure 2d). The tilted adsorption of **4** relates to the three-fold symmetric molecule being adsorbed on the four-fold symmetric NaCl surface, see Figure 3. The charge state and the spin state of **4** also have a subtle influence on the adsorption geometry, as will be discussed in the following (see also Figure S4).

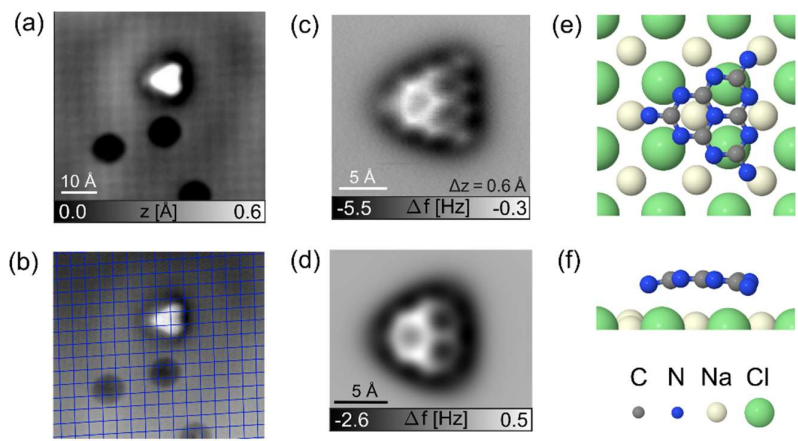

**Figure 3. Adsorption geometry of 4 on bilayer NaCl. (a)** Constant-current STM measurement of **4** at *I* = 1 pA, *V* = 0.2 V. A dark halo and a standing-wave pattern around the molecule indicate a negative charge state of **4**. The three dark circular features correspond to CO molecules. **(b)** Same data as (a) with a grid overlayed to extract the adsorption position of **4**. The grid's vertices indicate chlorine sites. **(c)** AFM measurement of **4** on bilayer NaCl. The tip-height offset Δ*z* from the setpoint *I* = 1 pA, *V* = 0.2 V is indicated. **(d)** AFM simulation based on the relaxed geometry of the anionic sextet **4** on NaCl. **(e)** Top-down view and **(f)** side view of the geometry-optimized on-surface adsorption position of anionic **4** on NaCl. The calculated molecular geometry is non-planar, with the nitrene's nitrogen atoms being close to sodium cations of the surface.

Our measurements indicate that **1**, **2**, and **3** are neutral on bilayer NaCl on Au(111), whereas **4** is negatively charged. Figure 3a shows an individual **4** exhibiting a dark halo around the molecule and a scattering pattern that is centered at the molecule's position. Such a pattern indicates interface-state electrons scattering at a charged adsorbate.[65,66] The dark halo can be attributed to a locally increased tunnelling barrier, due to a negative charge state.[65,67] In addition, Kelvin probe force spectroscopy data (Fig. S5) support a negative charge state of **4**. The precursor **1** and products **2** and **3** do not exhibit interface state scattering nor a dark halo indicating that they are charge-neutral on this surface. The charge state of **4** was stable within a bias window of *V* = ±0.7 V (see Fig. S3). When imaging **4** at a bias voltage of *V* = 0.8 V, it appears four-fold symmetric because of tip-induced changes of its adsorption orientation (see Fig. 4), caused by changes between four energy-degenerate adsorption orientations related to the four-fold symmetry of the NaCl surface, with small barriers for 90° rotations of the molecule.[68] This effect precluded imaging of the frontier orbital densities of **4**.

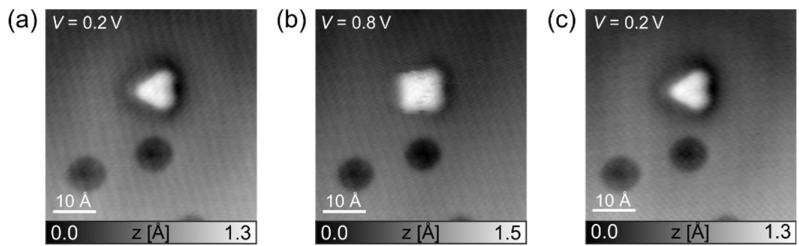

**Figure 4. STM measurements of 4 at different bias voltages. (a)** STM data, *I* = 0.75 pA, *V* = 0.2 V. **(b)** At *V* = 0.8 V, **4** appears four-fold symmetric. We interpret this by induced changes of the adsorption geometry, that is, changes between four energy-degenerate adsorption orientations, related to the four-fold symmetry of the NaCl surface, with small barriers for 90° rotations of the molecule.[68] These barriers might be overcome by inelastic electron tunneling processes or transient (dianionic) charging of **4**.[69] The effect precluded imaging of the frontier orbital densities of **4**. **(c)** Lowering the bias voltage to *V* = 0.2 V after image (b) was acquired, showed **4** with the same contrast as before, i.e., as in (a). The circular depressions correspond to CO molecules.

Complete active space perturbation theory (CASPT2) calculations (see Methods) of **4** give an adiabatic electron affinity of 5.29 eV. For Au(111), we assume a work function of $\phi_{Au(111)} \approx 5.26$ eV.[70] For bilayer

NaCl on Au(111) we expect a decrease of the work function by about 1 eV with respect to the bare Au(111) surface,[71,72] as is seen for thin alkali-halide films deposited on coinage metals.[72–77] Thus, the calculated electron affinity of **4** suggests an anionic charge state on bilayer NaCl on Au(111).[78] A neutral charge state of adsorbed **4** might be yielded by changing the substrate system toward one with a larger work function. However, this is not easily done by changing the metal substrate because Au(111) already features one of the largest work functions among all noble metal surfaces.

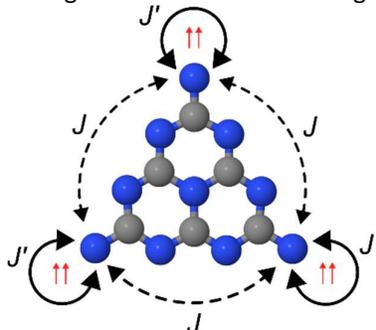

**Figure 5. Exchange coupling in trinitreno-*s*-heptazine, 4.** Considering the symmetry of the relaxed neutral **4** in the gas-phase, the spin-alignment description can be reduced to two values: intra-nitrene coupling *J'* (solid double arrows) and inter-nitrene coupling *J* (dashed double arrows). For each nitrene center, red arrows denote the relative spin orientation of unpaired electrons in the septet. Blue spheres denote nitrogen atoms, grey spheres denote carbon atoms.

The ground state of neutral **4** was computationally examined in the gas phase. Figure 5 schematically shows the exchange-coupling values within and between the three nitrene centers of **4**. Considering a threefold symmetry of **4** in gas phase, the inter-nitrene coupling *J* is assumed to be equal between all three nitrene centers, and intra-nitrene coupling *J'* to be the same for all three nitrene centers.

We obtain *J* and *J'* by mapping the broken-symmetry DFT (BS-DFT) and difference-dedicated configuration interaction (DDCI) results to a bilinear (BL) or bilinear-biquadratic (BLBQ) Heisenberg-Dirac-van Vleck (HDvV) spin Hamiltonian. Details on the calculations, spin Hamiltonians, and fitting procedure are given in Supplementary Note 2.

By mapping the broken-symmetry DFT results from the M06-2X hybrid functional calculations of **2**, **3** and **4** onto a BL spin Hamiltonian, we obtain the averaged intra-nitrene coupling of $J'(\mathbf{2,3,4})_{BS-BL}$ = 723.59 meV and, with averaged results of **3** and **4**, an inter-nitrene coupling of $J(\mathbf{3,4})_{BS-BL}$ = 5.80 meV. Mapping DDCI results of **2** onto a BL or BLBQ Hamiltonian, we obtain an intra-nitrene coupling of $J'(\mathbf{2})_{DDCI-BL}$ = 1092.06 meV and, with combined results of **3** and **4**, inter-nitrene couplings of $J(\mathbf{3,4})_{DDCI-BL}$ = 1.53 ± 0.06 meV and $J(\mathbf{3,4})_{DDCI-BLBQ}$ = 1.62 ± 0.07 meV for $B(\mathbf{3,4})_{DDCI-BLBQ}$ = 0.28 ± 0.13 meV.

All employed methods yielded positive values of *J* and *J'*, indicating ferromagnetic coupling between all six unpaired spins (septet ground state) for neutral **4** in the gas phase. All *J* and *J'* values for **2**, **3** and **4**, calculated using different levels of theory and their mapping onto different spin Hamiltonians, are noted in Tab. S7. The calculated *J'* values are in line with experimental intra-nitrene exchange coupling, e.g., about 650 – 870 meV for aryl nitrenes,[5] and about 1579 meV for NH nitrene.[8,79]

To compare the AFM measurements of **4** on NaCl with simulations, we conducted on-surface DFT calculations. For the calculations, we placed **4** on the NaCl surface according to the experimentally determined lateral adsorption position (Fig. 3b), and from this initial geometry relaxed the structures. Because the experiments indicate a negative charge state of **4** on NaCl/Au(111) (Fig. 3a), we allowed **4** to relax in the anionic (**4**$^-$) and dianionic (**4**$^{2-}$) charge states, considering all possible spin multiplicities for each charge state (see Methods). We performed AFM simulations on the relaxed geometries of all possible spin multiplicities for **4**$^-$ and **4**$^{2-}$ using the probe-particle model (see Fig. S4 and Methods).[59,80] The heptazine core is resolved in all AFM simulations. The AFM simulations (Fig. S4) show different contrasts of the heptazine core for different charge and spin states. We attribute these variations in contrast to differences in the DFT-calculated adsorption geometries. In our AFM simulations, we do not observe a brighter contrast at the nitrogen atoms compared to the carbon atoms, in contrast to the experiment (compare Fig. S4 and Fig. 3c,d) and full-density-based AFM simulations.[52,53] Furthermore, the probe-particle simulations show no atomic contrast at the nitrene centers. This could be explained by the nitrene-nitrogen atoms being closer to the substrate in the calculation than in the experiment. We find that the AFM simulations of the anionic doublet, i.e., **²4⁻**, and the anionic sextet, i.e., **⁶4⁻**, best match the experimental contrast of the heptazine core (see Fig. 2d and Fig. 3c), whereas the simulations for the dianion do not fit the experiment well, suggesting an anionic charge state of **4** on NaCl, in line with the calculated electron affinity of **4**. The DFT calculated energies indicate that the sextet is the anionic ground state. Thus, the combined results of calculations and experiment indicate that we experimentally observe **4** as an anionic sextet, i.e., **⁶4⁻**. A top-down view of the relaxed geometry of **⁶4⁻** is shown in Fig. 3e, a side view in Fig. 3f. The nitrene centers are located close to the sodium cations of the NaCl lattice, possibly because of electrostatic attraction between a sodium cation of the NaCl surface and the lone pair of each nitrene center.[52,53] The AFM simulation (see Methods) reproduces the experimental Δ$f$ contrast of the heptazine core and reflects the non-planarity of the molecule resulting from the adsorption site and orientation on NaCl (Fig. 3d).

Spin-density DFT calculations of **⁶4⁻** in the gas phase (see Fig. S5) suggest that the (six) spins on the nitrene centers are aligned ferromagnetically, similar to the septet ground state **⁷4⁰**, however, in the negative charge state of **4**, the additional electron is located at the heptazine core and is antiferromagnetically coupled to the unpaired nitrene electrons, resulting in the anionic sextet **⁶4⁻**.

**Conclusions:**
We have successfully generated mono-, di- and trinitreno-*s*-heptazine (**2**, **3**, **4**) from 2,5,8-triazido-*s*-heptazine (**1**) by tip-induced chemistry. By voltage pulses, the precursor's azide groups were dissociated, generating **2**, **3** and **4** on bilayer NaCl on Au(111). Theoretical investigations of neutral **4** using BS-DFT and DDCI indicate a ferromagnetic coupling between all three $S = 1$ nitrene centers, resulting in a high-spin, septet, ground state, i.e., **⁷4⁰**. Combined results of theory and AFM experiments indicate a sextet ground state of the anion of **4**, i.e. **⁶4⁻**, on bilayer NaCl on Au(111). Our study demonstrates that by tip-induced chemistry multiple nitrene centers can be successively obtained in a molecule, by dissociation of azide moieties, and how coupling between multiple nitrene centers can give rise to high-spin ground states. In future studies, on-surface-synthesized molecules with multiple nitrene centers could be used as building blocks for atomically precise superstructures with localized spin centers and for exploring possible coupling motifs between nitrene-containing molecules. Within a single molecule, the nitrogen-carbon core could be modified for the realization of different exchange couplings between individual nitrene centers.

**Author contributions:** L.-A.L., A.O. and L.Gross performed the experiments. L.-A.L. performed the DFT, BS-DFT calculations and AFM simulations. L.-A.L. and I.R. performed the CASSCF calculations and the calculation of exchange couplings. I.R. performed the DDCI calculations and programmed the DDCI-fitting routine. M.K. and F.E. synthesized the molecules. I.G. and L.Grill conceived the dissociation experiments. All authors discussed the results. L.-A.L. and L.Gross wrote the manuscript.

**Funding:** This work was financially supported by the H2020-MSCA-ITN ULTIMATE (grant number 813036), the European Research Council Synergy grant MolDAM (grant number 951519), and the European Research Council Advanced Grant AMOS (grant number 101097326).

**Notes:** The authors declare no competing financial interest.

**Supporting Information:** Additional experimental data measured on NaCl(2ML)/Au(111), description of the voltage pulses for nitrene generation, DFT and DDCI results and an overview of all calculated exchange coupling values.

**Methods:**

**Sample preparation, STM and AFM methods:**
The STM (scanning tunneling microscopy) and AFM (atomic force microscopy) measurements were performed in a home-built combined STM/AFM setup[37,81] operating at low temperature ($T \approx 5$ K) and under ultrahigh vacuum (UHV) conditions ($p \approx 1 \times 10^{-10}$ mbar). The mode of operation is described in refs. [37,82] The microscope is equipped with a qPlus sensor[83] operated in frequency-modulation mode[84] and the oscillation amplitude $A$ is kept constant at $A = 0.5$ Å. The tip is made from a 25 μm diameter PtIr-wire, sharpened *ex situ* by focused ion beam milling, and *in situ* by indenting the tip into the bare Au surface. A Au(111) single crystal was kept at 280 – 290 K while NaCl was evaporated to partially cover the Au(111) substrate with two monolayer (2 ML) thick NaCl(100) islands. 2,5,8-triazido-*s*-heptazine molecules (**1**) were sublimed onto the cold ($T < 15$ K) sample surface. In Figure S8 an overview STM image shows the prepared sample. For AFM imaging, a CO molecule was picked up by the tip from bilayer NaCl.[37,85] STM images were measured in constant-current mode with a sample voltage $V$ as indicated ($V$ is indicated as sample bias with respect to the tip at virtual ground potential). AFM images were acquired in constant-height mode at $V = 0$ V. The tip-height offset $\Delta z$ indicates the offset from the STM controlled setpoint. Positive $\Delta z$ values correspond to an increase in the tip-sample distance from the setpoint. Setpoint parameters for constant-current STM measurements and for AFM measurements are $I = 1$ pA, $V = 0.2$ V, unless indicated otherwise.

**On-surface DFT calculations:**
Density functional theory (DFT) calculations of **1** and **4** adsorbed on NaCl were performed within AiiDAlab[86] utilizing CP2K 9.1.[87] The PBE functional[88] with the Grimme-D3 dispersion correction was used.[89] The basis set that was used is TZV2P-MOLOPT-GTH. As a substrate, three layers of NaCl(100) were chosen, with the atom positions of the bottom layer being fixed, while the two top layers and the molecule were free to relax. The cell size in the sample plane is 6 × 6 NaCl unit cells with periodic boundary conditions. Planar neutral **1** and the planar anionic **4** were placed 2.7 Å above the topmost NaCl layer at the lateral adsorption position deduced from the experiment (see Fig. 3 and Fig. S1) and relaxed from that starting geometry. For **4**, geometry optimizations were conducted on the NaCl(100) surface with anionic (with $S = 1/2, 3/2, 5/2, 7/2$), and dianionic (with $S = 0, 1, 2, 3, 4$) charge states.

**Electron affinity calculations of trinitreno-*s*-heptazine:**
The adiabatic electron affinity of **4** was calculated using the fully internally contracted CASPT2 (complete active space second-order perturbation) approach as implemented in ORCA.[90] As a basis set cc-pVTZ (correlation-consistent polarized triple-zeta valence) was chosen. The active space ($n,m$) ($n$ electrons and $m$ orbitals) was chosen as (6,6) for the neutral septet and (7,6) for the anionic sextet. The gas-phase-optimized DFT geometry (B3LYP/def2-TZVP) of the neutral septet was used as the input geometry.

**AFM probe-particle model simulations:**
The probe-particle model[59] was used to simulate AFM images with the CO-tip preset with the following parameters: lateral resolution $dx = 0.1$ Å, oscillation frequency $f_0 = 30$ kHz, cantilever stiffness 1800 N/m, and oscillation amplitude $A = 0.5$ Å. The probe has a lateral stiffness of 0.25 N/m and radial stiffness of 30 N/m. Using a tip electrostatics model with $dz^2$-like charge density, the partial charge on the oxygen atom of the CO probe is -0.1 e.

**Broken symmetry DFT and configuration interaction calculations:**
The coupling values between spins on a nitrene spin-center (intra-nitrene coupling $J'$) and between two distinct nitrene centers (inter-nitrene coupling $J$) on a single molecule were calculated using broken-symmetry DFT and difference-dedicated configuration interaction (DDCI). All these calculations were carried out using ORCA.[90]

For the broken-symmetry DFT calculations, a geometry optimization of the structures of **2**, **3** and **4** has been performed for each high-spin configuration using the B3LYP[91] hybrid functional and def2-TZVP (triple-zeta valence orbitals with polarized functions) basis set. The Resolution of Identity approximation for Coulomb integrals (RI-J), with Chain-of-Sphere integration for Hartree-Fock exchange (COSX) was employed. The spins were then manually flipped on the respective nitrene center and the new electronic configuration was converged to a broken-symmetry solution. Keeping all else fixed, two different hybrid functionals with varying amounts of Hartree-Fock (HF) exchange were used: B3LYP (20% HF exchange)[91] and M06-2X (54% HF exchange).[92–94]

With DDCI($n,m$), mononitrene **2** was used to calculate intra-nitrene coupling $J'$, and **3** and **4** were used to calculate the inter-nitrene coupling $J$. For each DDCI($n,m$) calculation, a CASSCF($n,m$) calculation with unpaired nitrene electrons in the active space was used as the input. DDCI(2,2) was used for **2**, DDCI(4,4) for **3**, and DDCI(6,6) for **4**. The basis set for the DDCI calculations was cc-pVTZ, and only the low energy spectrum of magnetic excitations was probed for **2** and **3** (one singlet and triplet for **2**, one singlet, triplet, and quintet for **3**). The full spectrum was calculated for **4** (one singlet, three triplets, two quintets, and one septet).

**Table of Contents Graphic (TOC)**

Triazido-*s*-heptazine      Trinitreno-*s*-heptazine

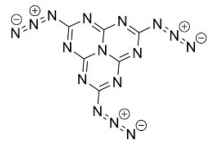 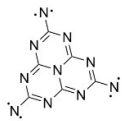

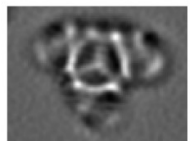 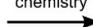 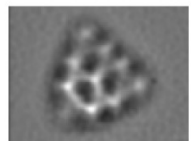

tip-induced chemistry

## Supplementary Information
## Tip-induced nitrene generation

Leonard-Alexander Lieske[1,*], Aaron H. Oechsle[1,#], Igor Rončević[2], Ilias Gazizullin[3], Florian Albrecht[1], Matthias Krinninger[4,†], Leonhard Grill[3], Friedrich Esch[4], Leo Gross[1,*]

[1] IBM Research Europe – Zurich, Rüschlikon, Switzerland
[2] Department of Chemistry, University of Manchester, Manchester, United Kingdom
[3] Physical Chemistry Department, University of Graz, Graz, Austria
[4] Chair of Physical Chemistry and Catalysis Research Center, Department of Chemistry, TUM School of Natural Sciences, Technical University of Munich, Garching, Germany

[#] Present address: Laboratory for X-ray Nanoscience and Technologies, Center for Photon Science, Paul Scherrer Institute, 5232-Villigen, Switzerland
[†] Present address: Leibniz Supercomputing Centre (LRZ) of the Bavarian Academy of Sciences and Humanities, Boltzmannstr. 1, 85748 Garching, Germany
*Corresponding authors, email: LAL@zurich.ibm.com, LGR@zurich.ibm.com

**Contents:**

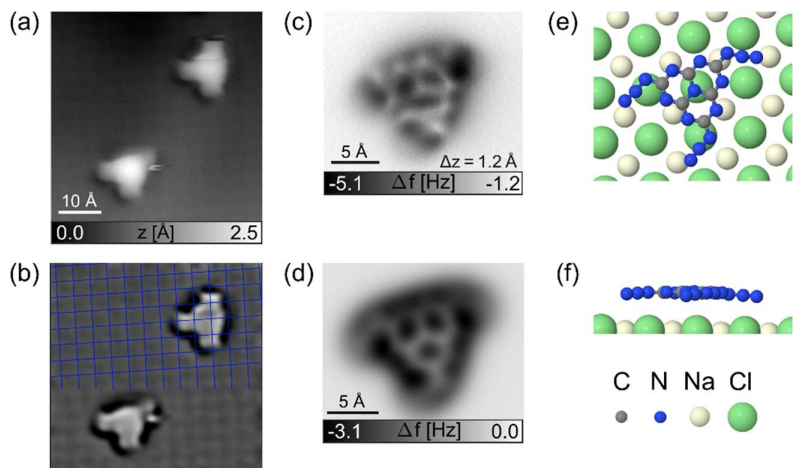

**Figure S1. Measured and calculated adsorption geometry of 1. (a)** Constant-current STM measurement of **1** at $I$ = 1 pA, $V$ = 0.2 V. **(b)** Laplace-filtered STM measurement from (a), a grid is fitted to the NaCl(100) lattice to extract the adsorption position of **1**. The grid's vertices indicate chlorine sites. **(c)** AFM measurement of **1** on bilayer NaCl. The tip-height offset $\Delta z$ from a setpoint of $I$ = 1 pA, $V$ = 0.2 V is indicated. **(d)** AFM simulation based on the relaxed geometry of **1** on NaCl. **(e)** Top-down view and **(f)** side view of the geometry-optimized on-surface adsorption of **1** on NaCl.

**Supplementary Note 1**
**Description of voltage pulses for nitrene generation**
Before a pulse was applied, the initial tip height was controlled by STM feedback with setpoint parameters $I_{SP}$ and voltage $V_{SP}$. Starting from this setpoint, the tip-height offset $\Delta z_P$ was applied. The voltage was linearly increased from $V_{SP}$ to the maximum applied voltage $V_P$ and back to $V_{SP}$ within the duration of the pulse of $t_P$. Typically, $t_P$ was 2 to 5 s. The current $I$ during the pulse was recorded; the maximum current during the pulse is $I_P$. An induced reaction or displacement of the adsorbate by the pulse typically resulted in a sudden drop of $I$ during the pulse. All voltage pulses were applied with the tip placed above a molecule. The voltage pulse parameters ($V_P$, and $\Delta z_P$) for nitrene generation by dissociation of the azide group, i.e., cleaving off $N_2$, were found iteratively.

The setpoint parameters $I_{SP}$ and $V_{SP}$ were chosen as either $I_{SP}$ = 1 pA, $V_{SP}$ = 0.2 V or $I_{SP}$ = 0.5 pA, $V_{SP}$ = 0.5 V. As initial parameters, $V_P$ = 2 V and $\Delta z_P$ = 1.5-2.5 Å were chosen, which resulted in $I_P$ < 1 pA and did not result in dissociation reactions. Using the same $V_P$, $\Delta z_P$ was reduced by about 0.1-0.2 Å in successive pulses until an azide group was dissociated or until $I_P$ reached values on the order of few tens of pA. If, at a given $V_P$ and with $I_P \gtrsim$ 10 pA, azide-group dissociation did not occur, we increased the $V_P$ by ~100 mV and increased $\Delta z_P$ to restore a small $I_P$. We repeated this sequence of $\Delta z_P$ adjustment and increase in $V_P$ until an azide group was dissociated. If an azide group was dissociated with a voltage pulse, its parameters were used as the starting point for the next azide-group dissociation from that molecule. The values of $I_{SP}$, $V_{SP}$ and $\Delta z_P$ determine the tip-sample distance and determine $I_P$ for an applied $V_P$. $I_P$ and $V_P$ values, at which an azide-group dissociation occurred, are stated for the voltage pulses in Fig. S3.

For specific tips and molecules, we observed that the sequential dissociation of azide groups from **1** to **2** to **3** in most instances required an increase of either $V_P$, $I_P$, or both for each subsequent dissociation step (see Fig. S3). The value of $V_P$ that was required for azide-group dissociation varied by ~100 mV for different tips and different molecules. We assume that the lateral position of the tip above the molecule could also influence $V_P$ and $I_P$. In our experiments, the molecules were usually also displaced on the surface and changed their orientation, when azide groups were dissociated. For that reason, we could not infer if selected azide groups could be dissociated, e.g., by a laterally off-centered tip position of the pulse. We did not observe azide-group dissociation when applying a voltage pulse with the tip positioned on the substrate near a molecule. Usually, pulses were applied with the tip centered above the molecule.

A detailed investigation of the mechanism of the azide-group dissociation was not the objective of this work, but a brief discussion is provided in the following. Typically, tip-induced voltage pulses result in cleavage of the weakest bonds at first.[1] Bond cleavage can be triggered by inelastic electron attachment[2] and might be mediated by surface state[3–5] or interface-state charge carriers,[1] electrons tunnelling resonantly into antibonding orbitals,[6] or Coulomb repulsion between parts of the molecule, due to charging.[7,8] Azide-group dissociation might be triggered by inelastic energy transfer from tunnelling electrons to the molecule by exciting vibrational modes.[2,9–13] In electron-mediated processes one or more electrons might be required to trigger a dissociation.[2,8,14] The electric field between tip and sample might contribute to the aforementioned mechanisms by modifying the reaction barrier for bond cleavage.[15–17]

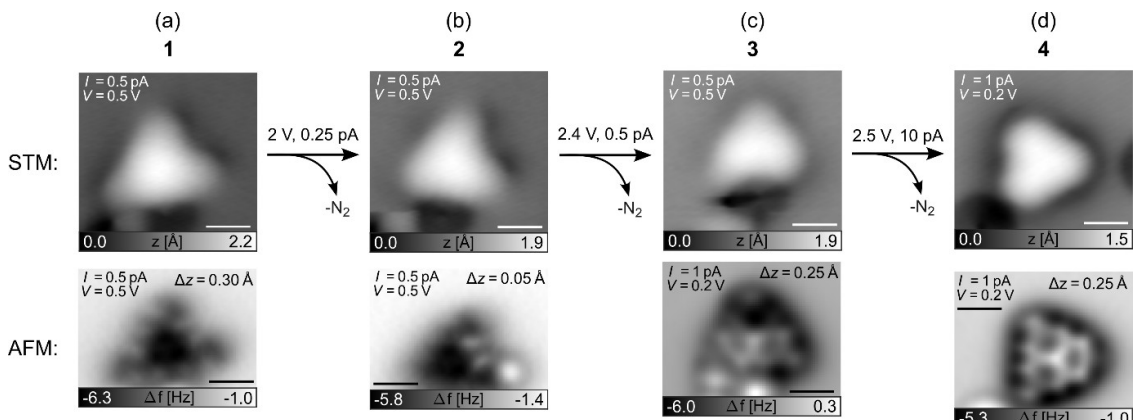

**Figure S2. Sequential generation of 2, 3 and 4 from 1.** Starting from **(a)** TAH **1**, **(b)** mononitrene **2**, **(c)** dinitrene **3** and **(d)** trinitrene **4** were generated by tip-induced voltage pulses. $V_P$ and $I_P$ are indicated for each azide-group dissociation. Constant-current STM measurements are shown in the first row, constant-height AFM measurements in the second row. Increased values of $V_P$ and $I_P$, indicated at the arrows, were needed for each subsequent azide-group dissociation on the reaction pathway from **1** to **4**. STM parameters, setpoints and tip-height offsets Δ$z$ are indicated in each panel. Scalebars are 5 Å.

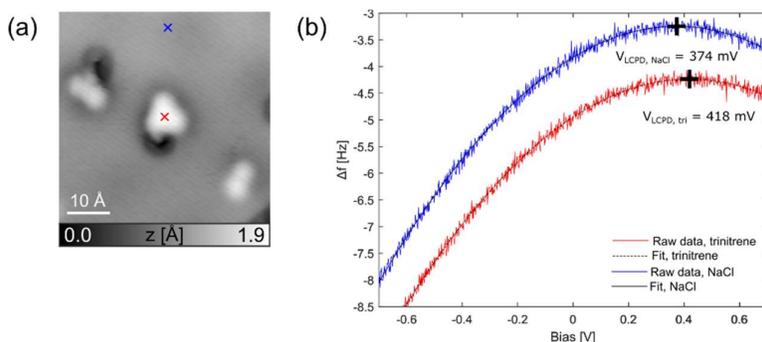

**Figure S3. Kelvin probe force spectra on 4. (a)** Constant-current STM measurement of **4**, $I$ = 0.5 pA, $V$ = 0.2 V. A CO molecule is located next to **4**. Crosses indicate where the Kelvin probe force spectra were recorded. **(b)** KPFS above **4** (red cross in (a)) and above the NaCl substrate (blue cross in (a)). Setpoint $I$ = 0.5 pA, $V$ = 0.5 V, Δ$z$ = 2 Å. Parabolas have been fitted to the measured Kelvin probe force spectra, and their peak positions are indicated. No charge bistability was observed in the measured bias range. The local contact potential difference (LCPD) shifts to a larger value comparing the measurement on **4** to the measurement on bare NaCl. This LCPD shift indicates a larger local work function on **4** in comparison to bare NaCl, in line with the assignment of **4** to a negative charge state (see Fig. 3, Fig. S2).[18,19]

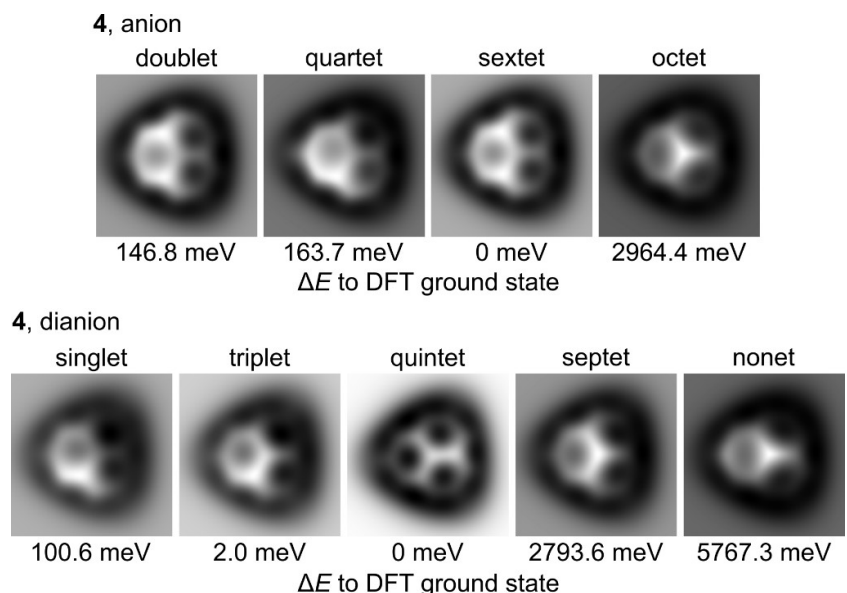

**Figure S4. PPM simulations of DFT results obtained for anionic and dianionic 4 in different spin states.** The experiment indicated a negative charge state of **4** on bilayer NaCl/Au(111). We considered a singly and doubly negative charge state and calculated both charge states in their respective possible spin multiplicities. An AFM simulation was run on each on-surface optimized geometry (see Methods for calculation details). The probe height is identical for all simulations. Of all performed AFM simulations, the anionic doublet and sextet configurations best match the AFM measurement of **4** (see Fig. 2d and Fig. 3c), indicating an anionic charge state. The sextet configuration has the lowest energy compared to the other spin multiplicities in the anionic charge state. The energies of the DFT-optimized geometries of possible spin multiplicities are indicated relative to the calculated spin ground-state of the respective charge state.

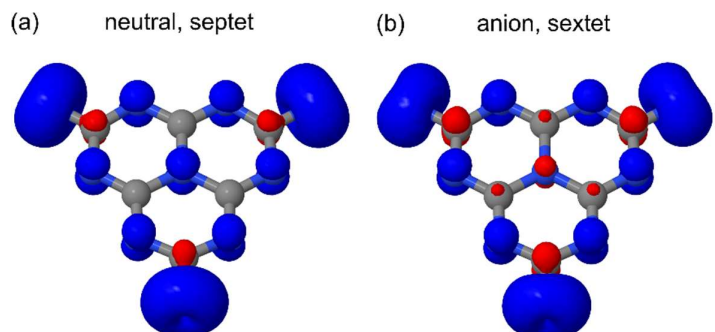

**Figure S5. Calculated spin density for neutral and anionic 4.** Gas-phase spin density of the **(a)** neutral, septet $^7\mathbf{4}^0$ and **(b)** anionic, sextet $^6\mathbf{4}^-$. Calculated using B3LYP/def2-TZVP in Psi4.[20] The spin-density isovalue is 0.01 e/$a_{Bohr}^3$. In (a) and (b) the spin densities are strongly localized on all three nitrene centers. Note that on the heptazine center, the calculated spin density of the anion is of opposite sign (minority spin) compared to the neutral molecule (majority spin).

**Supplementary Note 2**
**Additional computational results and discussion of the results**
**Broken-symmetry calculations and Heisenberg-Dirac-van Vleck spin Hamiltonian:**
$J_{BS}$ is obtained from the broken-symmetry DFT (BS-DFT) calculations by using Yamaguchi's equation:[21,22]

$$J_{BS} = \frac{E_{LS} - E_{HS}}{\langle S^2 \rangle_{HS} - \langle S^2 \rangle_{LS}}.$$

$E_{LS/HS}$ refers to the low-spin (broken-symmetry) or high-spin state's energy. $\langle S^2 \rangle_{HS/LS}$ is the total spin angular momentum of the high-spin/low-spin state. The six $\hat{S}_i = \pm 1/2$ spin centers (two per nitrene) can be described by a bilinear (BL) Heisenberg-Dirac-van Vleck (HDvV) Hamiltonian:[23]

$$\hat{H}^{BL} = -\sum_{\langle i,j \rangle} J_{ij}\, \hat{S}_i \cdot \hat{S}_j.$$

In this equation $J_{ij}$ is the exchange-coupling constant between spin operators $\hat{S}_i$ and $\hat{S}_j$ on sites $i$ and $j$. The threefold symmetry of **4** simplifies the assignment of $J_{BS}$ to $J$ by assuming $J$ to be equal between all three nitrene centers and $J'$ to be equal on all three nitrene centers. A positive (negative) coupling value indicates a ferromagnetic (antiferromagnetic) coupling between two spins.[24,25]

The intra-nitrene value $J'$ was calculated for **2**, **3** and **4**, and the inter-nitrene value $J$ for **3** and **4** (see Table S7). With def2-TZVP as the basis set, two different hybrid functionals were used to probe the influence of Hartree-Fock (HF) exchange on $J$ and $J'$.[26] B3LYP has a HF exchange of 20%,[27] M06-2X has 54%.[28] Calculations of coupling values for **2**, **3** and **4** for a hybrid functional consistently show similar values for $J$ and $J'$. The different amounts of HF exchange influence $J$ by less than a factor of two. For **3** and **4**, and $J'$ is of similar magnitude.

**Difference-dedicated configuration interaction and bilinear–biquadratic spin Hamiltonian:**
Broken-symmetry DFT calculations can exhibit spin-contamination in the low-symmetry (broken-symmetry) case,[22] and thus might provide inaccurate energies. To provide a comparative basis for the BS-DFT coupling values, difference-dedicated configuration interaction[29] (DDCI) was used. The DDCI-calculated excitation energies of **2**, **3**, and **4** are shown in Figure S6. These results were mapped to the BL HDvV Hamiltonian to obtain $J$ and $J'$.[30] We obtain $J'(\mathbf{2})_{\text{DDCI-BL}}$ = 1092.06 meV, $J(\mathbf{3})_{\text{DDCI-BL}}$ = 1.99 ± 0.23 meV and $J(\mathbf{4})_{\text{DDCI-BL}}$ = 1.49 ± 0.05 meV. From a combined fit of the results of **3** and **4** we obtain $J(\mathbf{3,4})_{\text{DDCI-BL}}$ = 1.53 ± 0.06 meV.

Because the BL HDvV model might not be best suited for systems with non-regular energy spacings[31] and for systems with three-spin exchange,[32,33] we also performed a non-linear fit of the DDCI energy-spectra using a bilinear–biquadratic (BLBQ) model. This model is a generalization of the bilinear HDvV Hamiltonian for $S = 1$ systems:[34,35]

$$\hat{H}^{BLBQ} = -\sum_i \left[ J \hat{S}_i \cdot \hat{S}_{i+1} + B (\hat{S}_i \cdot \hat{S}_{i+1})^2 \right].$$

The optimal values of $J$ and $B$ were determined by non-linear fitting of the DDCI results to the BL ($B$ = 0) or BLBQ Hamiltonians, and the errors were estimated from the covariance matrix as in ref [36]. Using this approach, the values of $J(\mathbf{3})_{\text{DDCI-BLBQ}}$ = 2.60 ± 0.07 meV; $B$ = 0.87 ± 0.08 meV and $J(\mathbf{4})_{\text{DDCI-BLBQ}}$ = 1.58 ± 0.04 meV; $B$ = 0.33 ± 0.09 meV. From a combined fit of **3** and **4** we obtain $J(\mathbf{3,4})_{\text{DDCI-BLBQ}}$ = 1.62 ± 0.07 meV; $B$ = 0.28 ± 0.13 meV. The difference between $J(\mathbf{3})_{\text{DDCI-BLBQ}}$ and $J(\mathbf{4})_{\text{DDCI-BLBQ}}$ might be explained by the presence of three-spin exchange in **4**.[32,33]

The relatively large intra-nitrene coupling $J'$, found by all used methods, can be attributed to the strong orbital overlap between the two ferromagnetically aligned spins on a nitrene center.[37] The relatively small inter-nitrene coupling $J$ between individual nitrene centers, found by all used methods, likely results from superexchange interaction mediated through the heptazine core.[38,39] The presence of heteroatoms in the heptazine core might hinder orbital delocalization weakening the inter-nitrene coupling.[40]

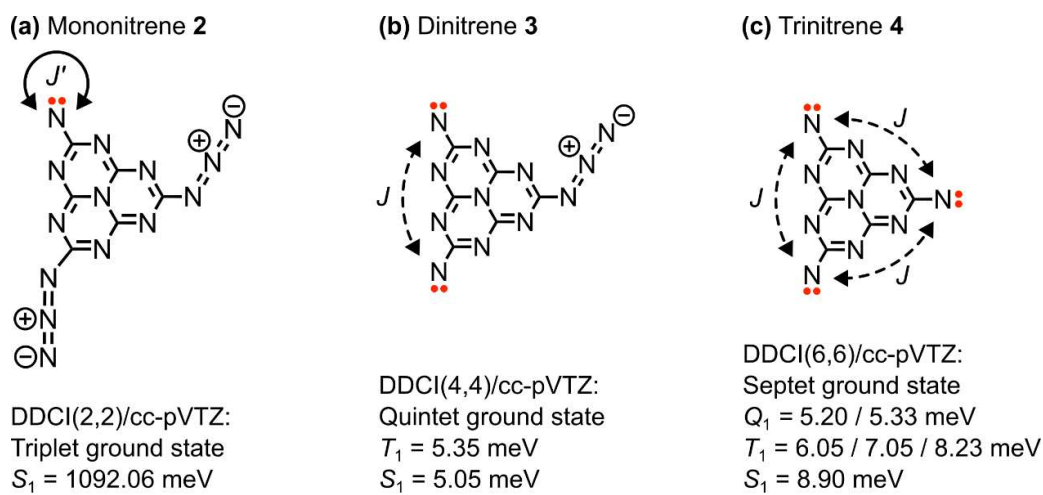

**Figure S6. DDCI-calculated excitation energies of 2, 3 and 4.** DDCI-calculated energy spectrum for **(a)** mononitrene **2**, **(b)** dinitrene **3** and **(c)** trinitrene **4**. The energy differences of the $S_1$ (singlet), $T_1$ (triplet) or $Q_1$ (quintet) states with respect to the ground state of each calculation are indicated. Multiple values for $Q_1$ and $T_1$ of **4** denote different roots. Red dots represent the unpaired electrons on the nitrene centers. For inter-nitrene coupling $J$ (dashed double arrow) and intra-nitrene couplings $J'$ (solid double arrow) see Tab. S7.

| Molecule | Level of theory | Inter-nitrene coupling ($J$) | Intra-nitrene coupling ($J'$) |
|---|---|---|---|
| Mononitrene **2** | BS-B3LYP/def2-TZVP (BL) | - | 446.49 meV |
|  | BS-M06-2X/def2-TZVP (BL) | - | 729.13 meV |
|  | DDCI(2,2)/cc-pVTZ (BL) | - | 1092.06 meV |
| Dinitrene **3** | BS-B3LYP/def2-TZVP (BL) | 6.26 meV | 445.77 meV |
|  | BS-M06-2X/def2-TZVP (BL) | 5.56 meV | 728.27 meV |
|  | DDCI(4,4)/cc-pVTZ (BL) | 1.99 meV | - |
|  | DDCI(4,4)/cc-pVTZ (BLBQ) | 2.60 meV | - |
| Trinitrene **4** | BS-B3LYP/def2-TZVP (BL) | 6.56 meV | 430.71 meV |
|  | BS-M06-2X/def2-TZVP (BL) | 6.04 meV | 713.36 meV |
|  | DDCI(6,6)/cc-pVTZ (BL) | 1.49 meV | - |
|  | DDCI(6,6)/cc-pVTZ (BLBQ) | 1.62 meV | - |

**Table S7. Exchange-coupling values $J$ and $J'$ by different levels of theory.** Coupling values $J$ and $J'$ for **2**, **3** and **4** were obtained from fitting the BS-DFT and DDCI calculation results to spin Hamiltonians (BL and BLBQ, see Supplementary Note 2). With def2-TZVP as the basis set, two different hybrid functionals were used to probe the influence of Hartree-Fock (HF) exchange on $J$ and $J'$.[26] B3LYP has a HF exchange of 20%,[27] M06-2X has 54%.[28] The DDCI-calculation results are displayed in Fig. S6, and all results are discussed in Supplementary Note 2.

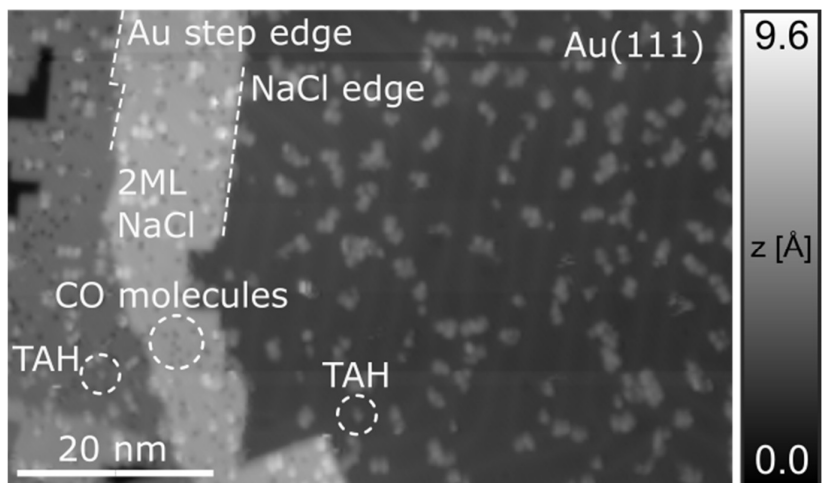

**Figure S8. STM overview image.** Overview of the sample surface showing the coverage of individual triazido-*s*-heptazine precursors (TAH, **1**) on a patch of bilayer NaCl(100) and on Au(111). Step edges, island edges, CO molecules and TAH **1** are indicated. Other molecules (not highlighted) are also present on the surface. Because of the high coverage, some molecules are grouped together. Parameters *I* = 0.5 pA, *V* = 0.5 V.